\documentclass[aps,prb,reprint,superscriptaddress,showpacs]{revtex4-1}
\usepackage{graphicx}
\usepackage{bm}

\begin{document}

\title{Ultracompact high-contrast magneto-optical disk resonator side-coupled to a plasmonic waveguide and switchable by an external magnetic field}
\author{Ji-Song Pae}
\affiliation{Department of Physics, Kim Il Sung University, Taesong District, 02-381-4410 Pyongyang, Democratic People's Republic of Korea}
\author{Song-Jin Im}
\email{ryongnam31@yahoo.com or sj.im@ryongnamsan.edu.kp}
\affiliation{Department of Physics, Kim Il Sung University, Taesong District, 02-381-4410 Pyongyang, Democratic People's Republic of Korea}
\author{Kum-Song Ho}
\affiliation{Department of Physics, Kim Il Sung University, Taesong District,  02-381-4410 Pyongyang, Democratic People's Republic of Korea}
\author{Chol-Song Ri}
\affiliation{Department of Physics, Kim Il Sung University, Taesong District,  02-381-4410 Pyongyang, Democratic People's Republic of Korea}
\author{Sok-Bong Ro}
\affiliation{Department of Physics, Kim Il Sung University, Taesong District,  02-381-4410 Pyongyang, Democratic People's Republic of Korea}
\author{Joachim Herrmann}
\email{jherrman@mbi-berlin.de}
\affiliation{Max-Born-Institute for Nonlinear Optics and Short Pulse Spectroscopy, Max-Born-Str. 2a,
D-12489 Berlin, Germany}
\date{\today}
\begin{abstract}
Here we propose and study a novel type of plasmonic resonators based on a metal-insulator-metal waveguide and a side-coupled magneto-optical disk controlled by an external magnetic field. The wavenumber change and the transmission of surface-plasmon-polaritons (SPPs) can be tuned by altering the magnetic field and reversible on/off switching of the running SPP modes by a reversal of the direction of the external magnetic field is demonstrated. Resonant enhancement of the magneto-plasmonic modulation by more than 200 times leads to a modulation contrast ratio more than tenfold ratio (90-\%-modulation) keeping a moderate insertion loss within an optical bandwidth of hundreds of GHz. Numerical simulations confirm the predictions by the derived analytical formulas of a high-contrast magneto-plasmonic modulation by the submicron ultra-small disk resonator.
\end{abstract}

\maketitle

Plasmonic waveguide-ring and disk resonators are important building blocks for many nanophotonic device components and key elements in the next-generation of integrated nanophotonic circuits. They provide several functionalities, e.g. filtering, wavelength division multiplexing and efficient modulators in integrated photonics. Localized surface plasmon modes in plasmonic resonators\cite{Yu2008,Cai2009,Hu2011,Im2016} have been intensively studied for the enhancement of electromagnetic fields. However, plasmonic cavities have a low quality factor on the level of several tens which is not enough to achieve a high contrast in the modulation of the plasmons. In contrast travelling resonant modes in waveguide ring resonators (WRR) have a higher quality factor. A WRR consists of a circular optical resonator with an unidirectional coupling mechanism to a straight waveguide. For certain wavelengths a WRR is in resonance when the waves interfere constructively in the ring introducing a significant phase shift near the resonance and resonantly enhance the interaction. To date various ring resonators based on plasmonic waveguides have been investigated such as on channel plasmonic waveguides \cite{Bozhevolnyi2006}, nano-plasmonic slot waveguides \cite{Zhu2011} or dielectric-loaded plasmonic waveguides \cite{Holmgaard2009}. Ultra-small cavity mode volumes can be achieved in plasmonic waveguide disk resonators as studied e.g. in \cite{Hao2007,Large2011,Zhai2015,Randhawa2011,Kuttge2010,Bo2017}.

Magnetoplasmonics using nanosystems with combined plasmonic and magnetic properties is a fast emerging new field and appealing for many applications. Including a magnetic material in a plasmonic structure provides a way of controlling plasmon propagation by using an external magnetic field (for a review see e.g. \cite{Armelles2013}). The magneto-optic effect allows switching that can be faster than those of other effects such as the electro-optic effect and thermo-optical effect. The magnetic field reversal by integrated electronic circuits can easily reach a speed in the GHz regime. Moreover, the recently demonstrated ultrafast magnetization switching with circularly polarized light enables a switching speed in the THz regime \cite{Stanciu2007}. However, the performance of the ultrafast magneto-plasmonic modulators in the visible and near-infrared range generally suffers from a small change of the SPP wavenumber \cite{Sepulveda2006,Khurgin2006} which requires a device size of at least tens of micrometers to achieve a sufficient interaction between the external magnetic field and SPP. The wavenumber change could be enhanced by introducing photonic crystals \cite{Yuu2008,Kuzmiak2012} where the transmission bandwidth is limited by photonic bandgaps. Plasmon-enhanced transverse magneto-optical Kerr effect \cite{Belotelov2011,Kreilkamp2013,Belotelov2013} was experimentally demonstrated to be resonantly sensitive to the magnetically-induced change of the SPP wavenumber and allow a 50-\%-modulation, but the corresponding structures are relatively bulky and not  fully waveguide-integrated. A more compact magneto-plasmonic interferometer using a hybrid ferromagnetic  structure was successfully demonstrated to present a 2-\%-modulation requiring a waveguide length of 20 $\mu$m \cite{Temnov2010,Temnov2016}, which can be improved to 12-\%-modulation \cite{Armelles2013} by adding a dielectric layer with higher refractive index. Magneto-optical properties of nanowires (see e.g. \cite{Melle2003} and references therein) and of the resonant inverse Faradayeffect in nanorings \cite{Joibari2014,Koshelev2015} has been recently studied.

In this paper we propose and study a novel approach for traveling SPP resonators by using a metal-insulator-metal waveguide (MIM) with a side-coupled magneto-optical disk (see Fig.1). In such ultra-compact structure the plasmon resonance enhancement is combined with the resonator resonance enhancement. Wavenumber and transmission of the travelling SPPs depend on the external magnetic field and on/off switching is demonstrated by the reversion of the direction of the magnetic field. We find a strong enhancement of the magneto-plasmonic modulation due to a high quality factor of the SPPs. The modulation contrast ratio reaches more than tenfold ratio (90-\%-modulation) keeping a moderate insertion loss within an optical bandwidth of hundreds of GHz. The fabrication of this kind of structure is possible within the current technology in this field. Recently, assisted by the rapidly developing nanotechnology, fabrication of circular or rectangular nanodisk cavity resonators have been reported, for examples (see e.g. \cite{Hao2007,Large2011,Zhai2015,Randhawa2011,Kuttge2010,Bo2017} and references therein).

\begin{figure}
\includegraphics[width=0.4\textwidth]{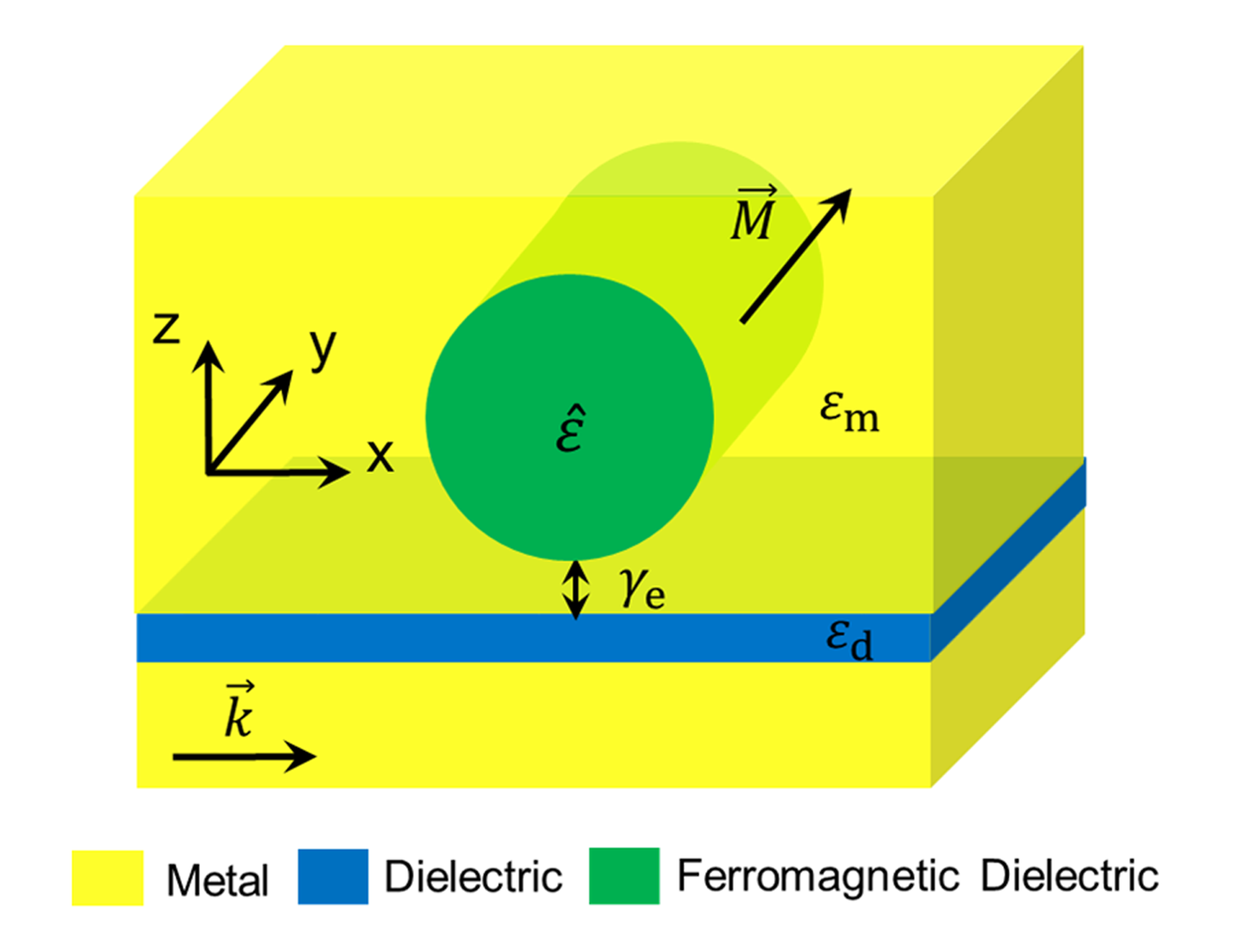}
\caption{The geometry of magneto-plasmonic disk resonator side-coupled to a metal-insulator-metal (MIM) waveguide.}
\label{fig:1}
\end{figure} 

\begin{figure}
\includegraphics[width=0.4\textwidth]{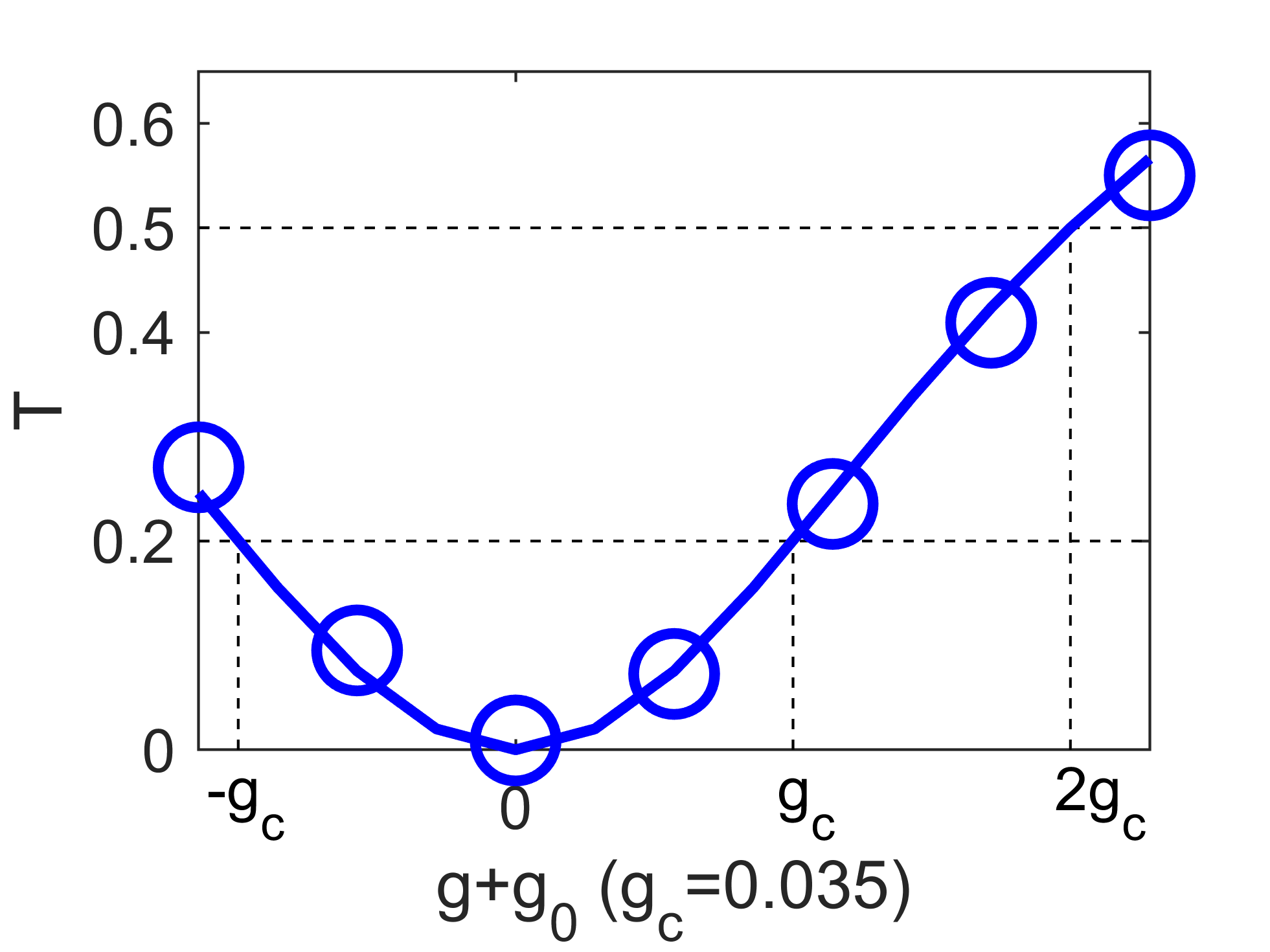}
\caption{Transmission $T$ of the MIM waveguide side-coupled to the disk resonator, as shown Fig. 1, in dependence of the gyration $g$ of the ferromagnetic filling of the disk resonator.  The blue circles present numerical results for a width of 50 nm of the waveguide, a radius of 400 nm of the disk and a distance of 20 nm between the waveguide and the disk resonator at the wavelength of 748 nm. The experimental data of the permittivity of silver \cite{Johnson1972} as ${\varepsilon _m}=-{\varepsilon _1} +i{\varepsilon _2}$ and the permittivity of Bi-substituted iron garnet (BIG) \cite{Dutta2017} as ${\varepsilon _3}$. ${\varepsilon _d}=2.25$ has been assumed. The blue curve is obtained by using the analytical formula Eq. (5), where ${g_c} = 0.035$  and ${g_0} = 0.03$ have been assumed to fit the numerical results.}
\label{fig:2}
\end{figure} 

\begin{figure}
\includegraphics[width=0.4\textwidth]{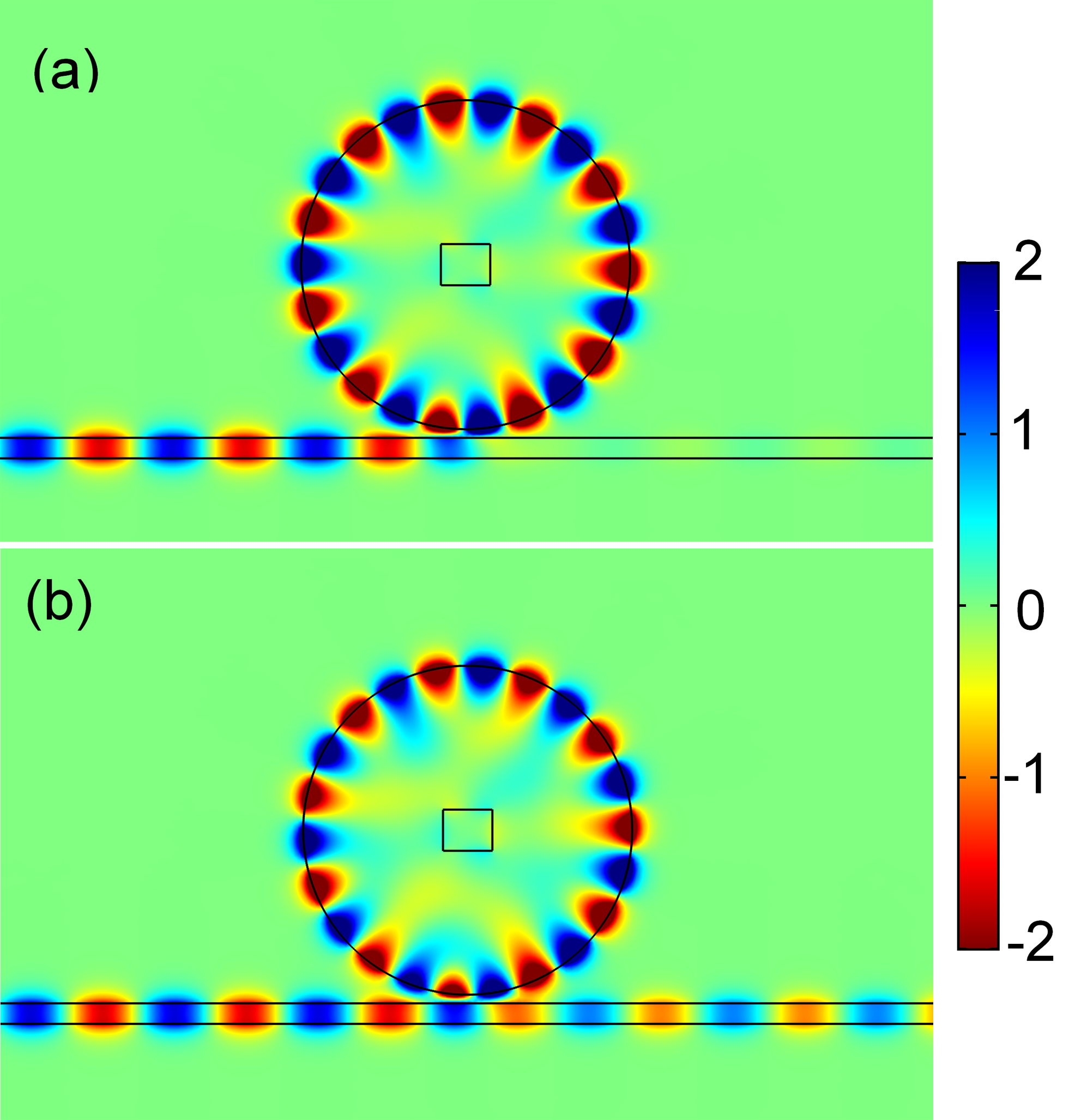}
\caption{Distributions of the magnetic field components of SPPs in the waveguide side-coupled to the disk resonator at a wavelength of 748 nm for $g =  - 0.03$ (a) and $g =   0.03$ (b), respectively. Other parameters are the same as in Fig. 2. The metal rectangle in the center of the disk is introduced to suppress the excitation of localized modes of the disk.}
\label{fig:3}
\end{figure} 

\begin{figure}
\includegraphics[width=0.4\textwidth]{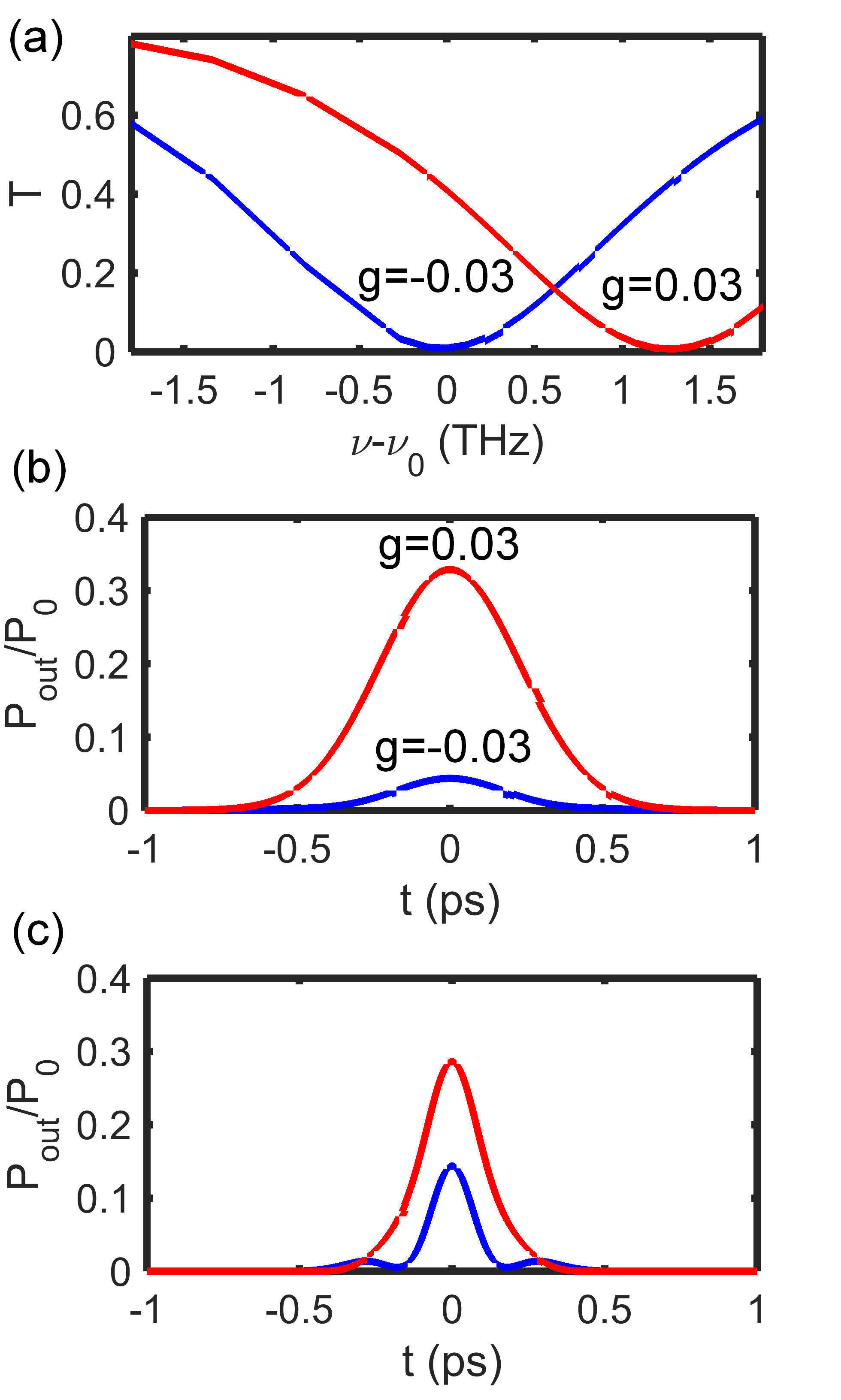}
\caption{Spectral and temporal responses of the magneto-plasmonic modulator. Transmission spectra according to a deviation of the frequency $\nu$ from the central frequency $\nu_0$ corresponding to a wavelength of 748 nm (a). The deviation of 1 THz corresponds to about 2 nm in wavelength units. Temporal power ${P_{\texttt{out}}}$ of the transmitted pulses normalized by a peak power ${P_0}$  of the input pulse (b,c). The full-widths at half maximum of the Gaussian input pulses are   500 fs (b) and 200 fs (c), respectively. The input pulses have a central wavelength of 748 nm. The blue and the red curves correspond to $g= -0.03$  and $g= 0.03$, respectively. Other parameters are the same as in Fig. 2.}
\label{fig:4}
\end{figure}

The geometry of a magneto-plasmonic disk resonator side-coupled to a MIM waveguide is shown in Fig. 1. Under an external magnetic field in the direction of y-axis, the permittivity tensor of ferromagnetic dielectric filling the disk resonator is expressed as

\begin{eqnarray}
\hat \varepsilon  = \left( {\begin{array}{*{20}{c}}
{{\varepsilon _3}}&0&{ig}\\
0&{{\varepsilon _3}}&0\\
{ - ig}&0&{{\varepsilon _3}}
\end{array}} \right).
\label{eq1}
\end{eqnarray}
Here the gyration $g$ in the off-diagonal components is proportional to the magnetization, $g = \beta M$, where $\beta$ is the magneto-optical susceptibility. The permittivity of the metal is ${\varepsilon _m}= - {\varepsilon _1} + i{\varepsilon _2}$, where ${\varepsilon _1} \gg {\varepsilon _2}$. For a sufficiently large disk radius, the wavenumber of the surface travelling magneto-plasmon approximates to that of the planar magneto-plasmon of a metal-ferromagnetic dielectric interface which is expressed as \cite{Im2017_1}
\begin{eqnarray}
\nonumber
k(g) = (1 - ag)k(0),\\
a = \frac{1}{{\sqrt {{\varepsilon _1}{\varepsilon _3}} (1 - \varepsilon _3^2/\varepsilon _1^2)}}.
\label{eq2}
\end{eqnarray}
The resonance condition for the SPP modes is determined by the condition $k(g)\cdot{2\pi r_{\texttt{eff}}} = 2\pi N$, where $N$ is an integer. The reonance frequency $\omega_0$ is sensitive to the disk radius and does not strongly depend on the other geometric parameters such as the waveguide thickness and the distance between the waveguide and the disk resonator. The resonance frequency $\omega_0$ is modulated by the external magnetic field as following.

\begin{eqnarray}
{\omega _0}(g) = \frac{1}{{1 - ag}}  {\omega _0}(0) \approx (1 + ag){\omega _0}(0)
\label{eq3}
\end{eqnarray}
The transmission of the SPP waveguide coupled to the disk resonator can be described by the temporal coupled-mode theory  \cite{Manolatou1999} from which the following results can be received:
\begin{eqnarray}
T(\omega ,g) = \frac{{{P_{\texttt{\texttt{out}}}}(\omega )}}{{{P_{\texttt{in}}}(\omega )}} = \frac{{{{[\omega  - {\omega _0}(g)]}^2} + {{({\gamma _e} - {\gamma _0})}^2}}}{{{{[\omega  - {\omega _0}(g)]}^2} + {{({\gamma _e} + {\gamma _0})}^2}}},
\label{eq4}
\end{eqnarray}
where $\omega$   and  ${\omega _0}$ are the considered angular frequency and the resonant angular frequency, respectively. ${P_{\texttt{in}}}$  and ${P_{\texttt{out}}}$  are the input and the output power of the SPPs, respectively. ${\gamma _0}$  and ${\gamma_{e}}$  are the decay rates due to the absorption in the resonator and the escape rate
due to the coupling between the waveguide and the resonator, respectively. If we assume the condition ${\gamma _e}/{\gamma _0} = 1$, which can be realized by adjusting the distance d between the disk resonator and the waveguide, at the frequency $\omega  = {\omega _0} - a{g_0}$  Eq. (3) and (4) lead to 
\begin{eqnarray}
T(\omega  = {\omega _0} - a{g_0},g) = 1 - \frac{1}{{{{[(g + {g_0})/2{g_c}]}^2} + 1}}.
\label{eq5}
\end{eqnarray}
Here, ${g_c} = 1/(2a{Q_0})$ is the characteristic gyration, where ${Q_0} = {\omega _0}/2{\gamma _0}$  is the quality factor of the surface travelling resonant mode of the unload disk resonator. Adjusting the magnetic field to the value $g =  \pm {g_0}$, Eq. (5) leads to 
\begin{eqnarray}
\label{eq6}
T(\omega  = {\omega _0} - a{g_0},g =  - {g_0}) = 0,\\
T(\omega  = {\omega _0} - a{g_0},g = {g_0}) = 1 - \frac{1}{{{{({g_0}/{g_c})}^2} + 1}}.
\end{eqnarray}
Eq.~(6) means the SPPs are completely dropped for these parameters. With reversed direction of the magnetic field the transmission is expressed as Eq.~(7).
From Eq.~(6) and (7), we can predict a high-contrast magneto-plasmonic modulation keeping a moderate insertion loss of $1/[{({g_0}/{g_c})^2} + 1]$ if the gyration by the external magnetic field  $g=g_0$ is on the level of the characteristic gyration ${g_c}$. Let us first make a simple order-of-magnitude estimation of the magnitude of the characteristic gyration ${g_c}$. The quality factor can be expressed as ${Q_0} = {\omega _0}({W_E} + {W_H})/(-d{W_E}/dt)$, where ${W_E}$  and  ${W_H}$ are energies of the electric field and the magnetic field in the disk resonator, respectively. Using the relations  $(d{W_E}/dt) =  - {\omega _0}{\varepsilon _0}{\mathop{\rm Im}\nolimits} (\varepsilon )\int {\left\langle {\vec E \cdot \vec E} \right\rangle dV}$, ${W_E} = (1/2){\varepsilon _0}{[d\omega \varepsilon (\omega )/d\omega ]_{\omega _0}}\int {\left\langle {\vec E \cdot \vec E} \right\rangle dV} $  and  ${W_H} = (1/2){\mu _0}\int{\left\langle {\vec H \cdot \vec H} \right\rangle dV}$ and substituting the electric and the magnetic field distributions of the planar metal-dielectric interface \cite{Maier2007}, we find that ${Q_0}$ is roughly estimated as ${Q_0} \approx \varepsilon _1^2/({\varepsilon _2}{\varepsilon _3})$ .With the experimental data of the permittivity of silver \cite{Johnson1972} as ${\varepsilon _1}$  and  ${\varepsilon _2}$  and the permittivity of Bi-substituted iron garnet (BIG) \cite{Dutta2017} as ${\varepsilon _3}$, at a wavelength of 750 nm we can estimate ${Q_0} \approx 280$. With $a \approx 0.06$ calculated by Eq. (2), the characteristic gyration ${g_c} = 1/(2a{Q_0})$  is about 0.03 which is in the order of magnitude of real ferromagnetic dielectrics \cite{Yao2015}. We note that the gyration value for BIG saturates at 0.06 for B=150 mT (see the Methods of Ref. \cite{Davoyan2014}). Obviously, the wavenumber modulation $a$ of surface travelling magneto-plasmons is enhanced by the high quality factor ${Q_0}$, allowing a small characteristic gyration $g_c$ which is experimentally achievable.
Fig.~2 shows the transmission $T$ of the SPPs in the waveguide side-coupled to the disk resonator as shown in Fig. 1 in dependence on the gyration $g$ of the ferromagnetic dielectric filling of the disk resonator. The blue circles have been obtained by numerical solutions of the Maxwell equation at the wavelength of 748 nm. The numerical results well agree with the prediction by the analytical formula Eq.~(5) (the blue curve), where   ${g_c} = 0.035$ and ${g_0} = 0.03$  have been assumed to fit the numerical results. Fig.~3(a) and (b) show the distributions of the magnetic field component of the SPPs at a gyration  $g =  - 0.03$ and $g= 0.03$ of Fig. 2, respectively. As seen by changing the direction of the magnetic field the transmission of the SPPs is switched from an off to an on state via the changed interference pattern.

Finally, let us consider the optical bandwidth of the proposed magneto-plasmonic disk resonator. Fig. 4(a) shows the transmission in dependence on the frequency obtained by the numerical solution of Maxwell equations in the frequency domain \cite{Im2016}. As seen a high contrast ratio of the transmission exceeding tenfold ratio (90-\%-modulation) is maintained within an optical bandwidth of hundreds of GHz. The temporal responses to input SPP pulses have been calculated numerically by the superposition of the responses to the Fourier components of the input pulses. In the case of an input pulse with a FWHM of 500 fs [Fig. 4(b)], one can see a tenfold-contrast ratio keeping a moderate insertion loss. In the case of a FWHM of 200 fs [Fig. 4(c)], a lower contrast and even distortion of the pulse shape is obtained, which are attributed to the reduction of the optical bandwidth of the resonant device.

In conclusion, we investigated a novel type of an ultracompact plasmonic resonator based on a MIM waveguide side-coupled to a magneto-optical disk resonator. Wavenumber and transmission of the SPPs of the disk resonator are modulated by the magnitude of the magnetic field and on/off switching is demonstrated by the change of the direction of the magnetic field. Moreover, the high quality factor of the surface travelling magneto-plasmon resonant modes allows strong enhancement of the maneto-plasmonic modulation by more than 200. A tenfold-contrast ratio of the magneto-plasmonic modulator with a submicron footprint is maintained within an optical bandwidth of hundreds of GHz. The analytical predictions have been well supported by the numerical results.

\bibliography{magdisk}
\end{document}